\documentclass[prd, aps, superscriptaddress, preprintnumbers, twocolumn, floatfix, nofootinbib]{revtex4}

\usepackage{amsfonts}
\usepackage{amsmath}
\usepackage{amssymb}
\usepackage{bm}
\usepackage{dcolumn}
\usepackage{epsfig}
\usepackage{graphicx}   
\usepackage{graphics}
\usepackage{color}
\usepackage[latin1]{inputenc}
\usepackage{latexsym}
\usepackage{rotating}
\usepackage{hyperref}

\usepackage{amsfonts}
\usepackage{amsmath}
\usepackage{amssymb}
\usepackage{bm}
\usepackage{dcolumn}
\usepackage{epsfig}
\usepackage{graphicx}
\usepackage{graphics}
\usepackage[latin1]{inputenc}
\usepackage{latexsym}
\usepackage{rotating}
\usepackage{hyperref}

\newcommand\be{\begin{equation}}
\newcommand\ba{\begin{eqnarray}}
\newcommand\ee{\end{equation}}
\newcommand\ea{\end{eqnarray}}

\begin{document}

\title{Dark Energy and Dark Matter in a Model of an Axion Coupled to a Non-Abelian Gauge Field}

\author{Stephon Alexander}
\email{stephonster@gmail.com}
\affiliation{Department of Physics, Brown University,
Providence, RI, 02912, USA} 

\author{Robert Brandenberger}
\email{rhb@physics.mcgill.ca}
\affiliation{Physics Department, McGill University, Montreal, QC, H3A 2T8, Canada, and \\Institute for Theoretical Studies,
ETH Z\"urich, CH-8092 Z\"urich, Switzerland}

\author{J\"urg Fr\"ohlich}
\email{juerg@phys.ethz.ch}
\affiliation{Institute of Theoretical Physics, ETH Z\"urich, CH-8093 Z\"urich, Switzerland}
\date{\today}

\begin{abstract}

We study cosmological field configurations (solutions) in a model in which the pseudo-scalar phase 
of a complex field couples to the Pontryagin density of a massive
non-abelian gauge field, in analogy to how the Peccei-Quinn axion field couples to
the $SU(3)$-color gauge field of QCD. Assuming that the self-interaction potential
of the complex scalar field has the typical {\it Mexican hat}
form, we find that the radial fluctuations of this field can act as {\it Dark Matter},
while its phase may give rise to tracking {\it Dark Energy}.
In our model, Dark-Energy domination will, however, not continue
for ever. A new component of dark matter, namely the one
originating from the gauge field, will dominate in the future.

\end{abstract}

\pacs{98.80.Cq}
\maketitle

\section{Introduction} 

Current observations \cite{Planck2015} show that about $95 \%$ of the energy in the
universe does not come from visible matter observed in ordinary laboratory experiments, but from
a new kind of matter in the form of {\it Dark Energy} and {\it Dark Matter}. Evidence 
for Dark Matter and Dark Energy comes exclusively from gravitational effects:
{\it Dark Matter} was first introduced 
to  account for the missing mass in galaxies \cite{Zwicky, Rubin} and galaxy clusters. Dark Matter
has the same gravitational interactions and produces the same gravitational effects
as regular matter in the form of a pressure-less gas, but it interacts only very weakly with visible matter and photons. The presence of Dark Matter is required in order to obtain the observed agreement between the angular power spectrum of cosmic microwave background (CMB) fluctuations and the power spectrum
of density fluctuations; (see, e.g., \cite{RHBrev} for a discussion of this point).
As compared to Dark Matter, far less is known about {\it Dark Energy}.
Its presence in the cosmos is required to explain the apparent accelerated
expansion of the Universe, as inferred from Supernova observations \cite{Perlmutter, Riess},
and to reconcile the spatial flatness of the Universe, as derived from CMB
anisotropy measurements \cite{Planck2015}, with the total energy density due to matter, including
Dark Matter, inferred from the observed dynamics of galaxies and galaxy clusters. 

In order to explain the data provided by these observations, the equation-of-state 
parameter $w$ of Dark Energy 
(namely the ratio of pressure to energy density) is now known to be close to $w = -1$.
Dark Energy could be due to a \textit{cosmological constant} in Einstein's field equation of the general theory of relativity. It could also be a manifestation of \textit{modified laws of gravity}, which become manifest only on cosmological scales. Or Dark Energy could be caused by
a \textit{new matter field} (``quintessence field'') with an unusual equation of state, $w \simeq -1$; (see, e.g.,
\cite{DEreviews} for recent reviews on the Dark Energy puzzle). In this paper
we focus our attention on the third scenario, which we call the
{\it quintessence} approach; (see \cite{quintessence} for some original references).

A fairly popular candidate \cite{axionDM} for Dark Matter is the {\it invisible axion}
\cite{axion, invisible}, a very light pseudo-scalar field originally introduced to solve the {\it strong CP problem}
of quantum chromodynamics (QCD) \cite{PQ}. This axion field couples linearly to the instanton (Pontryagin)
density of the $SU(3)$-color gauge field of QCD; (it plays the role of a dynamical vacuum angle). If the VEV of the axion field can be shown to vanish the strong CP problem of QCD is solved.

It has been postulated recently \cite{us} that {\it Dark Energy} could arise 
from another pseudo-scalar field, a new axion, that couples linearly to the Pontryagin density of a heavy
non-abelian gauge field operating at a high energy scale. The new axion could be conjugate to an anomalous current, 
$j^{\mu}_{\ell}$, that couples to the gauge field; see, e.g., \cite{Weinberg}. The chiral anomaly would then explain why the axion couples to the
Pontryagin density of the gauge field. (One might speculate that the anomalous current is leptonic and the gauge field is the weak $SU(2)$-gauge field.)

One of the challenges in the
{\it quintessence} approach is to explain why Dark
Energy is becoming dynamically important around the present time, and not
already in the very early universe, or in the remote future. If a cosmological constant were to be the source of Dark Energy we would be faced with the problem of explaining the precise, \textit{very small} value that
the cosmological constant would have to be given in order to explain the observational data. In our
{\it quintessence} approach to Dark Energy we want to avoid to be forced to
introduce a comparably tiny number by hand. {\it Tracking Quintessence} \cite{tracking}
is a way to cope with this problem. In models of tracking quintessence, the energy density of
the field responsible for Dark Energy follows the energy density of the dominant matter field
until times when a dynamical crossover prevents further decline of its energy density, and Dark Energy becomes the dominant form of energy in the Universe. In \cite{us} we have observed that the coupling of an
axion to the Pontryagin density of a massive non-Abelian gauge field can cause
slow rolling of the axion field, so that, as a consequence, the equation of
state of the axion field is the one required of Dark Energy, and this has yielded an interesting
scenario of tracking quintessence.

In this paper, we introduce a toy model of a complex field whose phase
(angular component) plays the role of a pseudo-scalar axion that is linearly coupled to the
instanton density of a massive non-abelian gauge field. This gauge field is invisible below rather high energy scales. Our model appears to describe, at once, Dark Matter and Dark Energy.
Both the radial and the angular components of the complex scalar field describe dynamical
degrees of freedom. While the radial component leads to Dark Matter, its phase
is a source of Dark Energy; (tracking quintessence).

If the coupling of the axion to the instanton density 
of the gauge field were neglected our model would yield a renomalizable quantum field theory, 
in contrast to the model studied in \cite{us}.

The organization of this paper is as follows: In the next section we introduce
our model and derive its field equations (of motion).
In Section 3 we discuss cosmological solutions of the classical field equations, assuming 
that the fields only depend on cosmological time. We show how
the radial component of the scalar field can play the role of Dark Matter, whereas
its phase is a candidate for Dark Energy, for reasons similar to those
advanced in \cite{us}. A discussion section concludes our paper.

The following notations and units will be used throughout: the cosmological scale factor is
denoted by $a(t)$, $z(t)$ is the associated cosmological redshift, and the 
Hubble expansion rate by $H(t)$; space-time
indices are denoted by Greek letters, group indices by latin letters; and we use natural units
in which the speed of light, $c$, and Planck's constant, $\hbar$, are set to $1$.

\section{The Model}

We consider a complex scalar field $\varphi$ with Lagrangian density
\be
{\cal L} \, = \, \frac{1}{2} \partial_{\mu} \overline\varphi \partial^{\mu} \varphi
- \frac{\lambda}{2} \bigl( | \varphi|^2 - R_0^2 \bigr)^2
- \frac{\mu^2}{2} |\varphi - {\bar{\varphi}}|^4,
\ee
where $\lambda$ and $\mu$ are dimensionless coupling
constants. This Lagrangian describes a renormalizable theory
\footnote{Note that fine-tuning of a mass term for the field $(\varphi - {\overline{\varphi}})$
to zero is assumed in Eq. (1). This renormalization condition is analogous to one appearing 
in the Coleman-Weinberg model \cite{CW}. In a follow-up work \cite{follow},
we will investigate ways to avoid this fine-tuning condition.}. For \mbox{$\mu^{2} = 0$,} 
the potential for $\varphi$ has the
usual ``Mexican hat" shape, with ground states
breaking the $U(1)$-symmetry of global phase transformations. The modulus of
the field minimizing the potential is denoted by $R_0$.
The term $\propto \mu^{2}$ breaks the $U(1)$-
symmetry explicitly; ($U(1)$- symmetry breaking is also encountered in the usual model of the Peccei-Qinn scalar in QCD).

We introduce polar coordinates in field space, i.e., radial and angular components
of $\varphi$,
\ba
\varphi \, &=& \, R e^{i \theta}, \\
R \, &=& \, R_0 + r \, , \nonumber 
\ea 
where $\theta$ is the angular component (the phase) of $\varphi$, $R$ its radial component,
and $r$ parametrizes radial fluctuations of $\varphi$ about a ground state configuration 
for $\mu^{2}=0$ corresponding to $\vert \varphi \vert = R_0$.

If the field $\varphi$ plays a role similar to the one the Peccei-Quinn scalar plays in
QCD then it must be coupled linearly to the Pontryagin density of some gauge field, $A_{\mu}^{a}$, 
which we here take to be a massive non-abelian gauge field
effective at a high energy scale. The coupling between the phase, $\theta$, of $\varphi$ and the gauge field $A_{\mu}^{a}$ is described by the following term in the Lagrangian
density of the theory 
\be \label{Pterm}
{\cal L}_P \, = \, - \alpha \theta F_{a \mu \nu} {\tilde F}_a^{\mu \nu}  \,.
\ee
Here $F_{a \mu \nu}$ is the field strength of $A_{\mu}^{a}$, 
$\mu$ and $\nu$ are space-time indices, while $a$ is a (gauge) group index, and 
$\alpha$ is a dimensionless coupling constant. 

Besides the field $\varphi$, we introduce an axial chemical potential $\mu_5$ conjugate 
to an axial vector current $J_{\mu}^5$ that couples to the gauge field $A_{\mu}^{a}$. The chiral anomaly is expressed by the equation
\be \label{anomeq}
\partial_{\mu} J_{\mu}^5 \, = \, \frac{2 {\tilde{\alpha}}}{\pi} {\rm tr} ({\vec{E}} \cdot {\vec{B}}) 
+ \text{terms} \propto \text{masses} \, ,
\ee
where $\tilde{\alpha}$ is a coupling constant. The axial chemical potential conjugate to $J_{\mu}^{5}$ appears in a term ${\cal L}_Q$ in the effective Lagrangian for the gauge field analogous
to (\ref{Pterm}), namely
\be \label{chiLag}
{\cal L}_Q \, = \, - {\tilde{\alpha}} \chi F_{a \mu \nu} {\tilde F}_a^{\mu \nu} \, ,
\ee
with
\be
{\dot \chi} \, = \, \mu_5 \, .
\ee
In the Appendix we discuss a possible origin of the 
(dimensionless) pseudo-scalar field $\chi$.

If all spatial gradient terms are neglected the Lagrangian for $\varphi$ becomes
\ba \label{sflag}
{\cal L} \, &=& \, \frac{1}{2} \bigl( {\dot r}^2 + R_0^2 {\dot \theta}^2 + 2 R_0 r {\dot \theta}^2 + r^2 {\dot \theta}^2 \bigr)
\nonumber \\
& & - \frac{\lambda}{2} \bigl( 2 R_o r + r^2 \bigr)^2 \\
& & - 8 \mu^2 \bigl( R_0^4 + 4R_0^3 r + 6 R_0^2 r^2 + 4 R_0 r^3 + r^4 \bigr) {\rm sin}^4\theta \nonumber \\
& & + \alpha\, \theta\, E \cdot B \, ,
\nonumber
\ea
where $E_{a}$ and $B_{a}$ are the electric and magnetic components of the field tensor of the gauge
field $A_{\mu}^{a}$, and
$$ E\cdot B:= \text{tr}(E\cdot B) = \sum_{i,a} E_{i}^{a}\,B_{i}^{a}.$$
We assume that this gauge field acquires a large mass at a phase
transition occuring at an early time denoted $t_m$.

In the following, the gauge group is chosen to be $SU(2)$. We make the following ansatz of a homogeneous gauge field configuration, expressed in terms of a scalar field 
$\psi(t)$ (see e.g. \cite{chromo}): 
\ba
A_{0}^{a}(t) \, &=& \, 0, \\
A_ {i}^{a}(t) \, &=& \, a(t) \psi(t) \delta_{i}^{a} \, ,
\ea
where $\delta_i^{a}$ is the Kronecker delta function. The ``electric field'' $E_{i}^{a}$ is then given by
\be
E_{i}^{a}(t) \, \sim \, a^{-1} (a \psi)^{\cdot} \delta_{i}^{a}
\ee
and the ``magnetic field'' by
\be
B_{i}^{a}(t) \, \sim \, g (a(t)\psi(t))^2 \, ,
\ee
where $g$ is the coupling constant of the non-abelian gauge theory.
In the following we will drop the group index. Note that the amplitude of the
magnetic field is suppressed, as compared to the one of the electric field, by 
the gauge coupling constant $g$ (which, later, we will take to be $g \ll 1$)
and by an additional factor of $a\psi$.

The field equations of motion for the components of the scalar field $\varphi(t)$ and
for $\psi(t)$ (or, equivalently, for the fields $\varphi$ and $E\cdot B$) describing 
a homogeneous and isotropic cosmology can be derived from the Lagrangian
(\ref{sflag}) to which the standard Yang-Mills Lagrangian for $A_{\mu}$ is added.
We are interested in solutions of the field equations describing small oscillations of $R$ about its ground state value $R = R_0$ and the 
response of the axion field $\theta$ to the gauge field $A_{\mu}$, as
determined by its coupling to the Pontryagin term tr($E\cdot B$). We thus linearize the field equations in
 $r/R_0$, and we will later assume that $\theta$ remains so small that we can approximate
${\rm sin}\theta$ by $\theta$. The radial equation of motion then becomes
\be \label{radialeq}
{\ddot r} + 3 H {\dot r} - R_0 {\dot \theta}^2 + 32 R_0^3 \mu^2 {\rm sin}^4\theta 
+ 4 \lambda R_0^2 r \, = \, 0 \, ,
\ee
and the angular equation is given by
\be \label{angulareq}
{\ddot \theta} + 3 H {\dot \theta} + 2 \frac{{\dot r}}{R_0}{\dot \theta} 
+ 32 \mu^2 R_0^2 {\rm sin}^3\theta {\rm cos}\theta \, = \,  8 \alpha R_0^{-2} E \cdot B \, ,
\ee
where we have kept the ${\dot r}/R_0$ term, since the term ${\dot r}/r$ will be
parametrically larger than $H$.

\section{Cosmological Solutions}

We follow the approach outlined in \cite{us} and search for solutions in which the terms 
$\propto \dot{\theta}, \ddot{\theta}$ in Eq. (\ref{angulareq})
can be neglected, so that this equation reduces to
\be \label{axioneq}
32 \mu^2 R_0^2 {\rm sin}^3\theta {\rm cos}\theta \, \simeq \,  8 \alpha R_0^{-2} E \cdot B \, .
\ee
Note that this relation between the axion field and the instanton density is identical to the one used in \cite{chromo, sorbo, shahin} 
to derive slow-rolling of an inflaton field at sub-Planckian field values.
In the small $\theta$ approximation Eq. \eqref{axioneq} reduces to
\be \label{thetavalue}
\theta^3 \, \simeq \, \frac{1}{4} \alpha \mu^{-2} R_0^{-4} E \cdot B \, .
\ee

The solutions of the radial field equation we are looking for describe small oscillations of $R$ 
about its ground state value, i.e., oscillations of $r$ about $0$. Assuming that $\theta$ is a solution 
of \eqref{thetavalue} and that $E\cdot B$ decays like an inverse power of time, the leading terms 
in the radial equation yield the equation
\be \label{radial}
{\ddot r} + 3 H {\dot r}  + 4 \lambda R_0^2 r \, = \, 0 \,,
\ee
which describes the motion of a damped harmonic oscillator. To solve \eqref{radial} we make the ansatz
\be \label{req}
r(t) \, \equiv \, x(t) e^{\sigma(t)} \,,
\ee
where $x$ and $\sigma$ are two real-valued functions of time $t$, with $\sigma$ chosen such that
terms $\propto \dot{x}$ in \eqref{radial} cancel. This requirement implies that
\be \label{sigmaeq}
{\dot \sigma} \, = \, - \frac{3}{2} H \, ,
\ee
The radial equation then reduces to
\be \label{hoscillator}
{\ddot x}  + \bigl( 4 \lambda R_0^2  - \frac{9}{4} H^2 - \frac{3}{2} {\dot{H}} \bigr) x \, = \, 0 \, .
\ee
Except at the beginning of the evolution of the Universe, the terms $\propto H^{2}$ and $\dot{H}$ 
in the frequency are negligible, and the solutions, $x(t)$, of \eqref{hoscillator} describe harmonic oscillations 
about $x=0$ with frequency, $\omega$, given by
\be
\omega \, = \, 2 \sqrt{\lambda} R_0 \, .
\ee
This is the mass of our dark matter candidate.
Note that, even at the beginning
of the evolution, this mass is larger than $H$ by a factor proportional to
$R_0 / m_{pl}$ (where $m_{pl}$ is the Planck mass), as follows from the Friedmann equation for $H$.

At this point we must verify that it is self-consistent to
neglect the terms in the original angular and radial equations of motion that we have omitted in \eqref{thetavalue} and \eqref{radial}. We first consider the angular equation of motion. We temporarily omit all coupling constants and factors of order unity from our equations.
The terms ${\ddot \theta}$
and $3 H {\dot \theta}$ are both of the order $O(H^2 \text{  } \theta)$. The terms, denoted $T$, we
have kept in the angular equation scale as
\be
T \, \sim \, R_0^2 \frac{E \cdot B}{\rho_R} \, ,
\ee
where $\rho_R \equiv R_0^4$, whereas
\be
H^2 \theta \, \sim \, H^2 \bigl( \frac{E \cdot B}{\rho_R} \bigr)^{1/3} \, .
\ee
At the initial time, $E \cdot B \sim \rho_R$. Thus, the ${\ddot \theta}$- and $3 H {\dot \theta}$-
terms are suppressed, initially, as compared
to the terms we have kept, by the square of the factor $H / R_0$. The Friedmann equations imply that this factor is of order $O(R_0 / m_{pl})$, which is expected to be tiny. Furthermore, as functions of time, the
terms ${\ddot \theta}$ and $3 H {\dot \theta}$ decay faster than $T$.
Hence, it is self-consistent to neglect the ${\ddot \theta}$- and $3 H {\dot \theta}$-
terms in Eq. \eqref{angulareq}.

In a similar way one may check that the term $2\, {\dot r}{\dot \theta} / R_0$ in \eqref{angulareq} is
negligible: Inserting the expression (\ref{thetavalue}) for $\theta$ into \eqref{angulareq}, comparison between
this term and the ones we have kept yields the condition
\be \label{constraint}
H r \, < \, R_0^2 \bigl( \frac{E \cdot B}{\rho_R} \bigr)^{2/3} 
\ee
for the term $\propto {\dot r}$ in \eqref{angulareq} to be negligible.
At the initial time, the left-hand side of \eqref{constraint} is suppressed, 
as compared to the right-hand side, by one
power of $R_0 / m_{pl}$. Neglecting the decrease in the amplitude of oscillation of $r$,
both terms would scale in the same way as a function of time. But since $r$
exhibits a damped oscillation, the left-hand side decreases faster in time than
the right-hand side.  Hence, our approximation \eqref{thetavalue} for the angular equation
\eqref{angulareq} is self-consistent. 

It is easy to see that neglecting the terms depending on $\theta$ in the radial equation \eqref{radialeq} is
self-consistent. We leave it to the reader to check this.

As will be shown in the following section, in the absence of any back-reaction of the scalar fields
(and/or other matter fields) on the gauge fields, one does not obtain a successful scenario for {\it tracking quintessence}: the energy density in $\theta$ will never increase relative to that in regular matter
and radiation. However, both the coupling of the scalar field to the Pontryagin density and the term proportional to the extra axial chemical potential that we have introduced in the Lagrangian 
affect the time evolution of the Pontryagn density ${\vec{E}} \cdot {\vec{B}}$ and make it decrease 
in time less rapidly than if those terms were absent.

The back-reaction of the scalar field on the gauge field can be
analyzed by following the analysis in our previous paper. In the
presence of the axion $\theta$ and of the chemical potential $\mu_5$,
the equation of motion for the electric field has a term proportional to
$(\alpha \dot{\theta} + {\tilde{\alpha}} \mu_5) B$,
\be \label{secular1}
{\dot E} + \kappa H E \, = \, - \bigl( \alpha {\dot \theta} + {\tilde{\alpha}} \mu_5 \bigr) B \, ,
\ee
where the constant $\kappa$ depends on whether the gauge field
has acquired a mass, or not, and whether we are in the radiation or
matter epochs. For a massive gauge field in the radiation era,
$\kappa = 3/2$, which, in the absence of coupling to the axion, i.e., for
$\alpha = 0$, leads to the scaling characteristic of matter
\be \label{correct}
E^2(t) \, \sim \, a(t)^{-3} \, .
\ee
The equations for $E$ and $B$ are equivalent to a second order differential
equation for the scalar function $\psi$, an equation given in \cite{chromo}
\footnote{Note that the nonlinear terms in the equations for $E$
and $B$ are suppressed for small values of $\psi$, i.e., at
late times.}.

Treating the right hand side of (\ref{secular1}) as a small 
perturbation, the solution of Eq. (\ref{secular1}), given initial
conditions at some time $t_i$, can be found in first-order 
Born approximation. It is given by 
\be \label{secular2}
E(t) \, = \, E_0(t) \bigl[ 1 + \int_{t_i}^t dt^{`} E_0(t^{`})^{-1} S(t^{`}) \bigr] \, \equiv \, E_0 + E_1 \, ,
\ee
where $E_0(t)$ is the solution describing a ``free'' $E$- field, and
\be
S(t^{`}) \, = \, \bigl( \alpha {\dot \theta}(t^{`}) + {\tilde{\alpha}} \mu_5 \bigr) B(t^{`}) \, .
\ee
The ``electric field'' $E(t)$ can also be written as
\be \label{secular3}
E(t) \, = \, E_0(t) \bigl[ 1 + G(t) \bigr] \, ,
\ee
where the factor $G(t)$ is called {\it secular growth factor}. 
This result also follows from the second order differential equation
for $\psi(t)$, see \cite{us},
\be
\psi(t) \, \sim \psi_0(t) \bigl[ 1 + G(t) \bigr] \, ,
\ee
where $\psi_0(t)$ is the solution found by setting $\alpha$ and ${\tilde \alpha}$ to $0$,
which corresponds to the field $E_0(t)$. Since $E$ is linear in $\psi$ and $B$ is
quadratic in $\psi$, both ``electric'' and ``magnetic'' fields acquire a
secular growth correction linear in $G(t)$, as long as $G(t) < 1$. The
secular growth term $E_1$ will begin to dominate over the background term
$E_0$ at some time $t_{sec}$ which we want to lie in the interval 
$t_{eq} < t_{sec} < t_0$. Once the
secular term starts to dominate over the background term, i.e., when $G(t) > 1$, the quantity $E^2$ scales
as $G^2(t)$, $ E \cdot B$ scales as $G^3$ and $B^2$ as $G^4$. We must
make sure that, at the present time, the quintessence field energy density,
which scales as $G^3$, dominates over the energy density of the new gauge field, which is
proportional to $E^2 + B^2$ (+ a term proportional to the mass of the gauge field) and scales as $G^4$. 
This will only be the case if the constant $\alpha$ is sufficiently large. 
(The order of magnitude of this constant will be discussed later).

The second term on the right hand side of (\ref{secular2}) leads to an 
extra contribution to $E \cdot B$. In our previous model this term had logarithmic
growth in time relative to the term present when the
coupling between scalar and gauge field is absent. Hence there will
be a time, denoted $t_{sec}$, when
the second term begins to dominate over the first, and we have
shown that, for $t \gg t_{sec}$, the new term can
come to dominate, yielding a {\it tracking Dark Energy} model.
A constraint on the viability of every such model is that
\be
t_{eq} < t_{sec} < t_0 \, ,
\ee
where $t_0$ is the present time.

In our present model the contribution to the source $S(t)$ 
originating from the axion decays too rapidly
in time to yield a significant growth factor. This is the reason 
why we have introduced the extra axial chemical potential $\mu_5$. 

Let us assume that $\mu_5$ is constant in time. Then
\be
\frac{E_1}{E_0} (t) \, \sim \, {\tilde{\alpha}} \mu_5 (t - t_i) \, ,
\ee
and the time, denoted $t_{sec}$, when $E_1$ starts to dominate and the
secular growth sets in is given by
\be \label{sectime}
t_{sec} \, \sim \, {\tilde{\alpha}}^{-1} \mu_5^{-1} \, .
\ee
As will be shown in the next section, a necessary condition for
a successful tracking dark energy scenario is
\be
t_{eq} \, < \, t_{sec} \, < \, t_0 \, .
\ee
Hence we must choose a tiny axial chemical potential $\mu_{5}$ satisfying
\be \label{crit}
t_{0}^{-1} \, < \, \tilde{\alpha} \mu_5 \, < \, t_{sec}^{-1} \, .
\ee
This condition could be met naturally 
if the axial chemical potential redshifts, as the
universe expands, until some late time, denoted by $t_{chem}$, with
$t_{chem} < t_{eq}$. An idea about a possible origin of
such a chemical potential is sketched in  Appendix A. 

\section{Cosmological Scenario}

We propose to interpret the radial component, $r$, of the field $\varphi$ 
as a component of {\it Dark Matter} and the
angular component, $\theta$, of $\varphi$ as describing {\it Dark Energy}. 
Since the potential for $r$ is close to a quadratic potential of a harmonic oscillator, 
$r$ performs damped oscillations about $r = 0$. This implies that its equation
of state is that of \textit{cold dark matter}, i.e., $w \equiv p / \rho = 0$, where $p$
and $\rho$ are pressure and energy density, respectively. Since
\be
\rho_r(t) \, \sim \, r^2(t) \, ,
\ee
it is easy to check from (\ref{sigmaeq}) and (\ref{req}) that, in the
radiation epoch as well as in the matter epoch,
\be
\rho_r(t) \, \sim \, a(t)^{-3} \, ,
\ee
which is the scaling that cold dark matter has. 

In the following, the contributions of the degrees of freedom described by the fields $r$
and $\theta$ to the total energy density of the Universe are studied.
We use the standard notation
\be
\Omega_X \, = \, \frac{\rho_X}{\rho_0} \, .
\ee
Here, $\Omega_X$ is the fraction of the total energy density the substance $X$
contributes to the total energy density, denoted by $\rho_0$, of a spatially flat
universe.

Since $\Omega_{DM}$, corresponding to Dark Matter, scales as
$\Omega_m$, corresponding to regular matter, for all times,
the coincidence condition that, at the present time, the magnitude of $\Omega_{DM}$ is 
comparable to the magnitude of the fraction, $\Omega_B$, of the total 
energy density contributed by baryons is a consequence of a similar 
condition assumed to hold at the time 
when the Standard Model matter fields acquire their mass.
This happens at a temperature $T \equiv T_{EW} \sim 250 {\rm GeV}$.

Assuming that the spontaneous breaking of our new
``Peccei-Quinn-like'' symmetry takes place at a temperature $T_m$ 
(corresponding to a time denoted by $t_m$) lower
than $T_{EW}$, the condition that will guarantee the right magnitude of the 
energy density of Dark Matter is given by
\be \label{OmegaReq}
\Omega_{DM}(T_m) \, \sim \, \frac{T_{eq}}{T_m} \, ,
\ee
assuming that the degrees of freedom described by the field $r$ do not decay into
other degrees of freedom, (an assumption whose validity is examined in Appendix B). 
In \eqref{OmegaReq}, $T_{eq} \sim 3 {\rm eV}$ is the temperature at the time,
$t_{eq}$, of equal matter and radiation when $\Omega_B \sim 1$.
The ratio in Eq. \eqref{OmegaReq} is of the order of $10^{-11}$ if $T_m \simeq T_{EW}$.
This initial condition can be realized if the number density of $r$ is
suppressed as compared to the number density of photons in
the same way as the baryon number.

The radial component $r$ starts to oscillate after the breaking
of the new symmetry. The temperature at which this symmetry
breaking occurs is
\be \label{Tcrit}
T_c \, \sim \, R_0 \, .
\ee
This can be inferred from the following argument: There are finite-temperature corrections 
to the potential of the scalar field, the leading such corrections being given by \cite{Dolan}
\be
\Delta V \, \sim \, g^2 T^2 |\varphi|^2 \, ,
\ee
where $g$ is a typical coupling constant. Invoking naturalness
we expect $g^2 \sim \lambda$; (see, e.g., \cite{RMP} for a review
of these arguments in the context of cosmology). In \eqref{Tcrit}, $T_c$
is the temperature where the negative contributions to the
quadratic term in the potential, expanded about $\varphi = 0$, cancel
the positive contribution from $\Delta V$. This implies (\ref{Tcrit}).
In the following, we will assume that $T_c = T_m$.

The energy density stored in the $r$ degrees of freedom at the time of the
phase transition is proportional to the potential energy density
before the phase transition, i.e.,
\be
\rho_{DM}(T_c) \, \sim \, \lambda R_0^4 \, ,
\ee
while the critical energy density is 
\be
\rho_{rad} \, \sim \, T_c^4 \, .
\ee
Hence
\be \label{OmegaPred}
\Omega_{DM}(T_c) \, \sim \, \lambda \, .
\ee
Comparing (\ref{OmegaReq}) with (\ref{OmegaPred}), we see that there
is a curve in the $(R_0, \lambda)$ parameter plane leading to the
Dark Matter density observed today \footnote{The possible decay
of quanta of the $r$- field into axions, i.e., the quanta of the $\theta$- field, is discussed in Appendix B.}.
Note that, as in the case of the QCD axion as a candidate for Dark Matter,
the mass of the field quanta of the $r$- field is not determined uniquely by the requirement 
that we obtain the Dark Matter density observed today. What is determined is a combination
of the coupling constant $\lambda$ and the Dark Matter particle mass, $\omega$, 
which must be given by
\be
\omega \, \sim \, \lambda^{-1/2} T_{eq} \, .
\ee

Next, we study the equation of state of the degrees of freedom described by the axion 
field $\theta$. Its potential energy density, $V$, is given by
\be \label{thetapot}
V(\theta) \, \simeq \, 8 \mu^2 R_0^4 \theta^4,
\ee
whereas the kinetic energy density, $K$, is given by
\be
K(\theta) \, \sim \, {\dot \theta}^2 H^2 R_0^2 \theta^2 \, .
\ee
Inserting the slow-roll solution (\ref{thetavalue}), one easily finds that 
\be
\frac{K}{V} \, \sim \, \bigl( \frac{R_0}{m_{pl}} \bigr)^{4/3} \, .
\ee
Thus, the equation of state parameter $w_{\theta}$ of the axion field $\theta$ is
\be
w_{\theta} \, \simeq \, -1,
\ee
and hence $\theta$ gives rise to {\it Dark Energy}.

It remains to show that, in fact, $\theta$ gives rise to {\it tracking Dark Energy},
i.e., its contribution to the total energy density $\rho_{0}(t)$ tracks the
contribution of the dominant component of matter until some late
time.

First, we infer from (\ref{thetapot}) and \eqref{thetavalue} that the energy density of $\theta$
(which is dominated by the \textit{potential} energy density) is given by
\be \label{rhotheta}
\rho_{\theta}(t) \, \simeq \, 2 \alpha \bigl( \frac{1}{4} \alpha \mu^{-2} \bigr)^{1/3} 
\bigl( \frac{E \cdot B}{\rho_R} \bigr)^{4/3} \rho_R \, .
\ee
At the time $t = t_m$ when the phase symmetry is broken we expect $\rho_{\theta}$ to be comparable
to $\rho_r$ (by equipartition of energy amongst the components
of the field $\varphi$), i.e.,
\be
\Omega_{\theta}(t_m) \,  \sim \, \frac{T_{eq}}{T_c} \, ,
\ee
(recall that $T_{c}\sim T_{m}$).

We propose to monitor the time evolution of $\Omega_{\theta}$. There are a couple
of key times. The earliest one is the time, $t_m$, when the gauge field becomes massive.
Before that time we have radiation scaling
\be \label{scaling1}
E \cdot B \, \sim \, a(t)^{-4}, \,\,\,\,\,\, t_i < t < t_m \, ,
\ee
while, afterwards, matter scaling prevails, i.e.,
\be \label{scaling2}
E \cdot B \, \sim \, a(t)^{-3}, \,\,\,\,\,\, t_m < t < t_{sec} \, .
\ee
The next later time of importance is the time, $t_{eq}$, of equal matter and radiation.
For $t < t_{eq}$, we have that $a(t) \sim t^{1/2}$, and, for $t>t_{eq}$, $a(t) \sim t^{2/3}$.
The third important time is the time when the secular growth of the electric field $E(t)$
becomes significant. We denote this time by $t_{sec}$. For $t > t_{sec}$, the
density $E \cdot B$ decreases less rapidly than $a(t)^{-3}$. The last important time
is the time, $t_{DE}$, after which Dark Energy dominates. We are assuming
here that
\be
t_i \, < \, t_m \, <  \, t_{eq} \,  <  \, t_{sec} \,  <  \, t_{DE} \, < \, t_0 \, ,
\ee
where $t_0$ is the present time.

From the equations derived above we can read off the scaling of $\rho_{\theta}$
in the various time intervals:
\be
\rho_{\theta} \, \sim \, t^{-8/3}, \,\,\,\,\ {\rm for} \,\,\,\,\, t_i < t < t_m \, .
\ee
Hence, during this first time interval, the quantity $\Omega_{\theta}$
is decreasing. This is the first phase in the evolution of the Universe after inflation.

During the second phase of evolution, we have that
\be
\rho_{\theta} \, \sim \, t^{-2}, \,\,\,\,\ {\rm for} \,\,\,\,\, t_m < t < t_{eq} \, .
\ee
This implies that $\Omega_{\theta}$ is constant, corresponding to tracking
behaviour of Dark Energy.

During the third phase of evolution, we again have that
\be
\rho_{\theta} \, \sim \, t^{-8/3}, \,\,\,\,\ {\rm for} \,\,\,\,\, t_{eq} < t < t_{sec} \, ,
\ee
which implies that $\Omega_{\theta}$ begins to
decrease again. However, after the time $t_{sec}$, the magnetic helicity $E \cdot B$ 
grows by an extra power of $G^3 \sim t^3$, and hence
\be
\rho_{\theta} \, \sim \, t^{+1/3}, \,\,\,\,\ {\rm for} \,\,\,\,\, t_{sec} < t \, ,
\ee
which implies that, after $t_{sec}$, $\Omega_{\theta}$ grows rapidly in $t$. 
Note, however, that the energy density of the new gauge field grows
even more rapidly, and we need to convince ourselves that it does not
dominate the total energy density before $V(\theta)$ has a chance to do so. The resulting
limits on the constant $\alpha$ will be discussed below.

Combining the initial condition for $\Omega_{\theta}$ at time
$t_m$ with the radiation scaling of $\rho_{\theta}$, for times $t > t_m$,
and the secular growth for times $t > t_{sec}$, we obtain
\ba \label{thetadens}
\Omega_{\theta}(t_0) \, & = & \, \Omega_{\theta}(t_{eq}) \frac{T_0}{T_{eq}} G(t_0)^3 \, \nonumber \\
& = & \, \frac{T_0}{T_c} G(t_0)^3 \, ,  
\ea
where $T_0$ is the present temperature of the cosmic microwave background,
$T_{eq}$ is the temperature at time $t_{eq}$, and $G(t_0)$ is the
secular growth factor between $t_{sec}$ and the present time, which
scales as $t_0^3$. This leads to a condition relating $t_{sec}$ to $T_c$,
\be
T_{sec} \, = \, T_0 \bigl( \frac{T_c}{T_0} \bigr)^{2/9},
\ee
for our scenario to explain why Dark Energy becomes dominant around
the present time $t_0$. 

As in most tracking quintessence models,
the coincidence problem of dark energy (why does dark energy
rear its head just at the present time) is not resolved.

The time evolution of $\Omega_{\theta}$ is sketched in Fig. 1.
The horizontal axis is time, the vertical axis is $\Omega_{\theta}$.
The figure shows that the energy density of the $\theta$- field can be interpreted as {\it tracking Dark Energy}. 

A final issue we must address concerns the size of the energy density,
\be \label{rhoA}
\rho_A(t) \, \sim \, E^2(t) + B^2(t) \, ,
\ee
carried by the gauge field. 

This energy density scales as matter, hence the gauge field $A_{\mu}$
makes a contribution to Dark Matter. We have to make sure that
$\rho_A$ remains negligible, as compared to $\rho_{\theta}$, until the present
time $t_0$. From time $t_m$ until time $t_{sec}$, when the secular term
begins to dominate, $\rho_A$ scales as matter, whereas $\rho_{theta}$
scales as radiation. A necessary
 condition on the viability of our scenario is
that, at the time $t_{eq}$ of equal matter and radiation, 
the energy density contributed by $\theta$ is larger than the one contributed by the gauge field, i.e.,
\be
\bigl( E^2 + B^2 \bigr)(t_{eq}) \, < \, \rho_{\theta}(t_{eq}) \, .
\ee
Making use of (\ref{rhotheta}) this leads to the condition that
\begin{equation*}
\bigl( E^2 + B^2 \bigr)(t_{eq}) 
\overset{!}{<} \,  2 \alpha \bigl( \frac{1}{4} \alpha \mu^{-2} \bigr)^{1/3}
\bigl( E \cdot B \bigr)(t_{eq})^{4/3} \rho_R^{-1/3} 
\end{equation*}
\be \label{upperbound}
 <   \, 2 \alpha \bigl( \frac{1}{4} \alpha \mu^{-2} \bigr)^{1/3}
\bigl( E^2 + B^2 \bigr)(t_{eq})^{4/3} \rho_R^{-1/3} 
\ee
which would guarantee that the energy density $\rho_A$ is smaller than $\rho_{\theta}$ 
at the time $t_{eq}$ of equal matter and radiation. The  upper bound on the right side of Eq. \eqref{upperbound}
follows from the Schwarz inequality. We can rewrite this condition as
\be \label{ccond1}
2 \alpha^4 \mu^{-2} \, > \, \frac{\rho_R}{(E^2 + B^2)(t_{eq})} \, .
\ee
Assuming that, at  time $t=t_m$, the energy density of the new gauge field
is comparable to $\rho_R$, condition (\ref{ccond1}) boils down to
\be \label{ccond2}
2 \alpha^4 \mu^{-2} \, > \, \bigl( \frac{z(t_m)}{z(t_{eq})} \bigr)^3 \, .
\ee
Since $V(\theta)$ grows as $G^3$, for $t>t_{sec}$, whereas $E^2 + B^2$ 
increases in $t$ as $G^4$, a necessary (but not sufficient!) condition
on the coupling constant $\alpha$ needed to ensure that our candidate for quintessence
dominates over the energy density of the new gauge field \textit{at the present time} $t_0$
is that there is an additional secular growth factor $G$ multiplying the right side of (\ref{ccond2}), 
for times $t$ between $t_{sec}$ and the present time.
Using that $t_{sec} = t_{eq}$ and that $G \sim z^{-1}$, a reasonable estimate for today's value of $G$ is
given by $G \sim 10^4$. This
does not change the condition (\ref{ccond2}) on $\alpha$ by more than
one order of magnitude. Taking the temperature at $t_m$ to be
$250 {\rm MeV}$, and taking $\mu \sim 1$ we find that
 $\alpha > 10^6$. This represents quite a
severe fine-tuning requirement, which is, however, \textit{much less severe} than the fine-tuning
that would have to be imposed on the bare cosmological constant.
 
This is a condition involving the values of the constants
$\alpha$ and $\mu$ and the initial energy density of the
gauge field and can be satisfied as long as $\alpha$ is
large and $\mu$ is small.

Once the secular term starts to dominate (i.e. $G(t) > 1$), 
the energy density $\rho_{\theta}$ increases as $g G(t)^3$ (where
we recall that $g \ll 1$ is the gauge coupling constant). Initially the
energy density $\rho_{A}$ scales as $G^2$ since the contribution
from the magnetic field (which scales as $g^4 G^4$) is suppressed
compared to the contribution from the electric field which contains
no power of $g$. Eventually - roughly speaking when 
\be
g^2 G^2 \, \sim \, 1 \, ,
\ee
the magnetic field contribution catches up, and from then on
$\rho_A$ increases more rapidly than $\rho_{\theta}$. Thus,
our model predicts that the phase of dark energy domination comes
to an end at some point in the future.

\begin{figure*}[t]
\begin{center}
\includegraphics[scale=0.8]{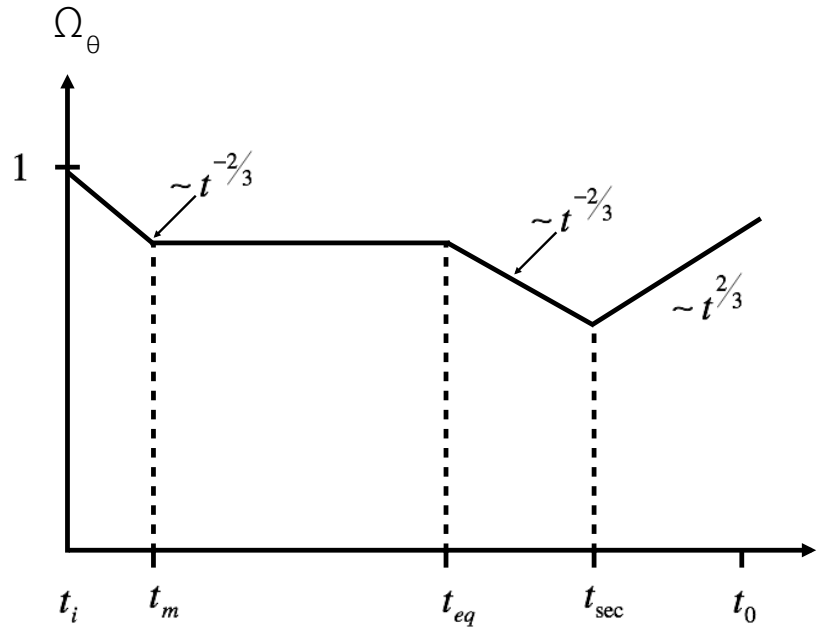}
\caption{Sketch of the time evolution of the fractional contribution $\Omega_{\theta}$
of the $\theta$ field to energy density of the Universe. The horizontal axis is time,
the vertical axis is the value of $\Omega_{\theta}$. The value of $\Omega_{\theta}$
initially decays from $t_i$ until $t_m$. Between $t_m$ and $t_{eq}$
we have exact tracking, i.e. $\Omega_{\theta}$ is constant. For
$t > t_{eq}$ the value of $\Omega_{\theta}$ initially decreases until
the timie $t_{sec}$ when the secular growth term for the electric field $E$
becomes important, after which $\Omega_{\theta}$ grows (for illustrative purposes
we have chosen $G(t)$ such that the growth is proportional to the scale factor).}
\end{center}
\end{figure*}

\section{Conclusions and Discussion}

We have proposed a model involving a complex scalar field $\varphi$ that can give rise
to both {\it Dark Matter} and {\it Dark Energy}. {\it Dark Matter} 
is provided by the radial oscillations of the field $\varphi$ about its symmetry breaking
minimum, {\it Dark Energy} by the angular variable, which is a new axion. A key feature of our 
model is a coupling of the axion to the Pontryagin density of a
non-abelian gauge field. The field $\varphi$ is introduced in analogy to
the Peccei-Quinn scalar of QCD.
The phase of $\varphi$ couples to the Pontryagin density of the gauge field. This provides
a mechanism for very slow rolling of the angular variable $\theta$, so that $\theta$
can yield {\it Dark Energy}. In turn, the dynamics of $\theta$, assisted by
an additional axial chemical potential, induces
secular growth of the electric component, $E$, of the gauge field. 
Once the secular growth term in $E$ starts to dominate over the usual term, 
the contribution of $\theta$ to the total energy density starts to grow.
Thus, $\theta$ is a candidate for {\it tracking quintessence}.

In our model, the energy density, $\rho_A$, of the gauge field
represents an extra contribution to Dark Matter. For
sufficiently large values of the coefficient $\alpha$ one can ensure
that $\rho_A$ is negligible at the present
time. However, eventually $\rho_A$ will grow faster than the density
of Dark Energy. Thus, our model predicts that the period
of Dark Energy domination does not continue arbitrarily far into
the future.
 
In our setup, the approximate equality of the energy densities in
{\it Dark Matter} and {\it Dark Energy} has a natural explanation
since the energy densities of the two components are proportional
during most of the evolution of the universe (from $t_m$ until $t_{eq}$).
For $t_i < t_m$ and for $t_{eq} < t_{sec}$ the contribution of
$\theta$ decays relative to that of {\it Dark Matter}, whereas
it increases after $t_{sec}$.We need $t_{sec}$
to lie in the interval $[t_{eq}, t_0]$.

If $\theta$ is to be a viable candidate for {\it Dark Energy}, it has
to be very weakly coupled to electromagnetism \cite{Carroll}.
This is why we need to introduce a new gauge field which
$\varphi$ couples to. Since, in our setup, {\it Dark Matter} and {\it Dark Energy}
belong to the same sector, our model predicts that {\it Dark Matter}
has negligible interactions with regular matter. Direct detection of
{\it Dark Matter} in accelerator experiments or in underground 
laboratories would rule out our scenario.

In our model, {\it Dark Matter} is coupled to {\it Dark Energy}. This
coupling gives rise to interesting predictions on observations, as
was studied in toy models of the two dark sectors in \cite{Abdalla}
and references therein. Work on this topic is in progress.

As for the QCD axion, we have to cope with a
potential domain wall problem \cite{DW}. If the values of the potential
at field values $\theta = 0$ and $\theta = \pi$ are
exactly the same, then if the $\varphi$ field
begins in thermal equilibrium and undergoes a symmetry
breaking phase transition a network of domain walls
will inevitably form by causality \cite{Kibble}. 
This network would acquire a ``scaling solution'' (the network looks the
same at all times when lengths are scaled to the Hubble
radius $t$) and would persist to the present time. A single
domain wall in our Hubble radius would overclose the
universe if the symmetry breaking scale is above roughly
$1 {\rm TeV}$; (see e.g. \cite{CSrevs} for reviews of
the cosmology of topological defects). We can avoid this
domain wall problem in the same way it is avoided for
QCD axions. For example, we could slightly lift the
potential to make $\theta = 0$ the unique vacuum state.
We could also assume that an early period of
cosmological inflation provides the causal connections
on super-Hubble scales which leads $\varphi$ to
fall into the same vacuum state everywhere in the
observable part of the Universe.

There has been other recent work connecting the two
dark sectors in the context of QCD-like theories; see, e.g., \cite{Stephon1} 
(which is based on \cite{Stephon2}). 

\section*{Acknowledgement}
\noindent
One of us (RB) wishes to thank the Institute for Theoretical Studies of the ETH
Zurich for kind hospitality. RB acknowledges financial support from the ``Dr. Max
R\"ossler-'' and the  ``Walter Haefner Foundation'', and from the ``ETH Zurich Foundation'', 
as well as through a Simons Foundation fellowship. His research is also supported in
part by funds from NSERC and the Canada Research Chair program.

\section*{Appendix A: Origin of the Axial Chemical Potential}

In this section we present a possible scenario for the origin of the
axial chemical potential $\mu_5$. Let us consider a second scalar field
$\chi$ coupling to $E \cdot B$, in analogy to the angular field variable $\theta$,
i.e., with a coupling given by (\ref{chiLag}). We take $\chi$ to have vanishing
mass dimension, as assumed in (\ref{chiLag}). We can introduce a scalar
field ${\tilde{\chi}}$ with the usual mass dimension $1$ by setting
\be
{\tilde{\chi}} \, \equiv \, \chi_0 \chi \, ,
\ee
where $\chi_0$ is some mass scale.

Let us assume that ${\tilde{\chi}}$ has an exponential potential of the form
\be
V_{{\tilde{\chi}}} \, = \, V_0 \bigl[ e^{- {\tilde{\chi}} / \chi_0} - 1 \bigr] \, ,
\ee
where the constant $V_0$ has mass dimension four. We also assume
that ${\tilde{\chi}}$ couples to some heat bath. This induces a correction
to the effective potential whose leading term is (see e.g. \cite{Dolan} and
the review in \cite{RMP})
\be
\delta V \, = \, \frac{1}{2} T^2 {\tilde{\chi}}^2 \, .
\ee

If we assume that ${\tilde{\chi}}$ tracks the minimum of the
effective potential we find that 
\be
\chi \, \sim \, \frac{V_0}{\chi_0^2} \bigl( \frac{T_{eq}}{T} \bigr)^2 T_{eq}^{-2} \, ,
\ee
for ${\tilde{\chi}} \ll \chi_0$, which leads to
\be
{\do{\chi}} \, \sim \frac{4}{3} \frac{V_0}{\chi_0^2} \frac{1}{t_{eq}} T_{eq}^{-2}\, ,
\ee
for $t < t_{eq}$, and
\be
{\do{\chi}} \, \sim \frac{4}{3} \frac{V_0}{\chi_0^2} \frac{1}{t_{eq}} 
\bigl( \frac{t}{t_{eq}} \bigr)^{1/3} T_{eq}^{-2}\, ,
\ee
for $t > t_{eq}$. Making use of the Friedmann equation to
express the time $t_{eq}$ in terms of the energy density
$T_{eq}^4$ at that time, we find that
\be \label{expr}
{\dot{\chi}} \, \sim \, \frac{V_0}{\chi_0^2 m_{pl}} \, .
\ee
The fact that there
is a factor of $m_{pl}$ (= Planck mass) in the denominator of (\ref{expr})
makes it possible to obtain a small value of ${\dot{\chi}}$,
which leads to a small value of the axial chemical potential
$\mu_5$, as required by the criterion (\ref{crit}). It does take
some tuning of $V_0$ and $\chi_0$ to obtain a value of
$\mu_5$ which lies exactly in the range given by (\ref{crit}).

\section*{Appendix B: Possible Resonance Effects}

In our scenario, the radial field $r$ is oscillating. It is
coupled to the angular variable $\theta$ via the nonlinear
terms in the equations of motion. We must hence
worry about possible resonance effects like the
parametric instability by which the oscillations of
the inflaton field at the end of the period of inflation
induce exponential growth of fields coupled to the
inflaton \cite{TB, DK} (see also \cite{RRevs} for recent
review articles).

To study the possible resonant excitation of $\theta$
due to the oscillations of $r$ we consider the equation
of motion for fluctuations of $\theta$ about the background
value $\theta_0$ considered in the main text:
\be
\theta \, = \, \theta_0 + \theta_1 \, .
\ee
In the small amplitude limit for the fluctuation $\theta_1$ we
have
\be
{\ddot \theta_1} + 3 H {\dot \theta_1} + \frac{{\dot r}}{R_0} {\dot \theta_1} 
\, = \, 0 \, .
\ee
This is in fact a first order differential equation for $\chi \equiv {\dot \theta_1}$
which has the solution
\be
{\rm ln} \bigl( \frac{\chi}{\chi_i} \bigr) \, = \, - 3 {\rm ln} \frac{t}{t_i}
- \frac{2}{R_0} \int_{t_i}^t dt' {\dot r}(t') \, ,
\ee
where $t_i$ is the initial time and $\chi_i$ is the value of $\chi$ at that time.
Since the integrand on the right hand side of the above equation
is oscillating, there is clearly no resonant growth.

Above, we have shown that oscillations of the $r$ field
does not induce a parametric resonance instability for
$\theta$ fluctuations. However, 
to ensure that our estimate of the dark matter density
from the $r$ field is correct, we must also ensure
that the perturbative decay of $r$ is not too efficient. For
an interaction Lagrangian describing the decay of a canonically
normalized field $r$ into another canonically normalized field
$\chi$ \footnote{The coefficient $g$ here
has nothing to do with the gauge coupling constant in the main
part of the text.}
\be
{\cal L}_{int} \, = \, - g \sigma r \chi^2 \, ,
\ee
the perturbative decay $\Gamma$ rate is given by
\be
\Gamma \, = \, \frac{g^2 \sigma^2}{8 \pi m} \, ,
\ee
where $m$ is the mass of the oscillating field. In our
case $\chi = R_0 \theta_1$, where $\theta_1$
is the fluctuation of $\theta$ about the slow-roll
solution $\theta$ given by (\ref{thetavalue}). 
Expanding the Lagrangian (\ref{sflag}) to leading
quadrtic order in $\theta_1$ we can read off
what corresponds to $g \sigma$ in the general
case. The mass of $r$ can be read off of the
same Lagrangian. Then, one finds that
in our case the ratio
\be
\frac{\Gamma}{H} \, \sim \, \mu^{5/3} \lambda^{-1/2} \alpha^{2/3} 
\bigl( \frac{E \cdot B}{R_0^4} \bigr)^{2/3} \, ,
\ee
and we can see that it is not hard to choose parameters
and initial conditions on the energy density in the new
gauge field such that already at the time $t_m$
\be
\frac{\Gamma(t_m)}{H(t_m)} \, \ll \, 1 \, ,
\ee
and that hence perturbative decay is
negligible. Hence, our dark matter candidate does not
decay efficiently into dark energy.

\end{document}